# Critical Analysis of an FeP Empirical Potential Employed to Study Fracture of Metallic Glasses

Yezeng He,[1,2] Peng Yi[2,5] and Michael L. Falk[2-5*]

[1]*School of Materials Science and Engineering, China University of Mining and Technology, Xuzhou, 221116, People's Republic of China*

[2]*Materials Science and Engineering, Johns Hopkins University, Baltimore, Maryland 21218, USA*

[3]*Mechanical Engineering, Johns Hopkins University, Baltimore, Maryland 21218, USA*

[4]*Physics and Astronomy, Johns Hopkins University, Baltimore, Maryland 21218, USA*

[5]*Hopkins Extreme Materials Institute, Johns Hopkins University, Baltimore, Maryland 21218, USA*

**Abstract**

An empirical potential that has been widely used to perform molecular dynamics studies on the fracture behavior of FeP metallic glasses is shown to exhibit spinodal decomposition in the composition range commonly studied. The phosphorous segregation induces a transition from ductility to brittleness. During brittle fracture the atomically sharp crack tip propagates along a percolating path with higher P concentration. This embrittlement is observed to occur over a wide range of chemical compositions, and toughness decreases linearly with the degree of compositional segregation over the entire the regime studied. Stable glass forming alloys that can be quenched at low quench rates do not, as a rule, exhibit such thermodynamically unstable behavior near to or above their glass transition temperatures. Hence, the microstructures exhibited in these simulations are unlikely to reflect the actual microstructures or fracture behaviors of the glassy alloys they seek to elucidate.

*Corresponding author.
E-mail address: mfalk@jhu.edu (M.L. Falk).

# 1. Introduction

Metallic glasses (MGs) have drawn considerable attention as promising structural materials due to their unique combination of properties such as high strength, extreme hardness and superior corrosion resistance [1-5]. However, the limited ductility of MGs restricts their use in various structural, engineering and functional applications [6]. Considerable scientific efforts have been made to unravel the physics of deformation and fracture in these materials in order to guide efforts to increase their toughness. Even so, controversies remain about the failure and fracture mechanisms of MGs [7,8].

Recent theoretical and experimental studies have demonstrated that many factors will influence the fracture behavior of MGs. For instance, Lewandowski et. al. [9] compared properties of numerous MGs and reported a critical Poisson's ratio exists around 0.31~0.32 below which the MGs are observed to be brittle. However, no fundamental explanation for this empirically observed relation currently exists. Murali et al.[10] has investigated the susceptibility of $Zr_{41.2}Ti_{13.75}Cu_{12.5}Ni_{10}Be_{22.5}$ MGs to embrittlement upon annealing and found that the toughness declines by up to 90%. Rycroft and Bouchbinder [11] attributed the annealing-induced embrittlement transition to the existence of an elastoplastic instability that induces cavitation ahead of the crack tip for sufficiently relaxed glasses. This is supported by recent experimental studies that link the fictive temperature, sometimes also referred to as the effective temperature, to a brittle-to-ductile transition in these materials that is higly sensitive to loading rate [12]. This reinforces the relevance of a number of simulation studies of cavitation in MGs that have been undertaken with the stated aim of more fully understanding the fracture process zone [13-15]. Moreover, the residual stress, and alloy composition may also play important roles in the fracture behavior of MGs [16-18]. Despite intense investigation, theoretically predicting the fracture behavior of MGs remains a challenge.

In recent years, molecular dynamics (MD) simulations have been used to directly examine brittle and ductile fracture behavior in MG materials [19,20]. A number of embedded atom method (EAM) interatomic potentials have been used for these studies including a common model for CuZr [21] and an FeP potential that was originally created to mimic the behavior of P impurities in steels [22]. Notably the FeP potential appears to be one of the few models of an amorphous alloy that results in brittle behavior on the sub-micron scale, making it an attractive system for examining a range of mechanical response regimes using atomistic simulations. However, it is important to note that this potential was not designed to mimic realistic phase behavior in either the crystalline or liquid phases. The main motivation for the development of this potential was described upon its publication as such, "For molecular dynamics purposes in studying reactor steels, the interesting region is that of small concentrations ($\sim 10^{-3}$) of P in Fe, in particular, the behaviour of point defects in lattices." The authors of this potential further note that, "Pure P is covalently bonded and cannot be described by this type of potential. We therefore do not attempt to fit properties of pure phosphorus, or phosphorus-rich compounds, concentrating instead on point defects in α-iron…" Therefore, we ask the questions, "What is the underlying phase behavior of this potential?" and "How do the thermodynamic driving forces arising from this potential influence the fracture behavior exhibited by this model?" Our investigation provides evidence that this particular FeP potential exhibits spinodal decomposition, and it is the resulting high degree of phase separation, which depends sensitively on the thermal processing history, that determines the fracture behavior of the simulated MG model.

## 2. Methods

In this study, MD simulations were used to investigate the fracture behavior using the EAM potential parametrized by Ackland et al. [22] that has been previously applied to model FeP MG fracture. [19,20] All the simulations were conducted using LAMMPS software package [23]. The Nosé-Hoover thermostat and Parinello-Rahman barostat are used to control the temperature and pressure, respectively. The velocity-verlet algorithm is used to integrate the equations of motion with a time step of 2 fs. Periodic boundary conditions were applied in all directions. Two different ways were used to prepare a system, and we will refer to these as melted-and-quenched (MQ) and annealed (A). For MQ systems, 162,000 atoms were initially arranged as a disordered BCC structure with P atomic percentage of 15%, 20% or 25%. The temperature of this system was raised gradually to 2000 K and equilibrated for 20 ns, followed by a cooling process to 1 K at a quench rate of 1 K/ps and a subsequent relaxation process for 2 ns at 1 K. Alternatively, the systems were prepared through annealing as follows. Initially consisting of 162,000 Fe atoms, the systems were first processed as the melted-and-quenched simulation as explained above. Then a fraction of Fe atoms (15%, 20% or 25%) were randomly substituted with P atoms. Subsequently, we anneal these systems for 1ns at 1K, 500K, 600K, 700K and 800K. We designate these systems A1, A500, A600, A700 and A800, respectively. For the above simulations, the NPT ensemble was employed and the pressure was maintained at zero. The resulting samples have dimensions of approximately 530Å×265Å×15Å. To study the fracture phenomena, we create a fracture specimen from each sample by introducing a sharp crack of length 100 Å through removing atoms in the middle of the sample. Finally, the samples were stretched at a strain rate of 0.05/ns along the direction perpendicular to the crack face. During the tensile process, the temperature was maintained at 1K to eliminate the thermal effects using the Nosé-Hoover thermostat and the boundaries perpendicular to the crack face were held fixed to eliminate the Poisson effects.

## 3. Results and discussion

To start with, we consider the $Fe_{85}P_{15}$ system as an example to illustrate the structure of system obtained via the traditional melting-and-quenching process. In Figure 1(a), compared to the initial configuration, the final configuration of a MQ sample shows P aggregation, in which the P atoms aggregate into a continuous fractal structure. To verify that this system is thermodynamically driven toward P aggregation, we consider the thermodynamic relation:

$$\Delta G_{mix} = -T\Delta S_{mix} + \Delta H_{mix}, \qquad (1)$$

Where $\Delta G_{mix}$, $\Delta S_{mix}$ and $\Delta H_{mix}$ are the change of mixing Gibbs free energy, entropy and enthalpy, respectively. The mixing entropy can be calculated using an ideal solution approximation, which is valid in the dilute limit,

$$\Delta S_{mix} = -R(x_{Fe} \ln x_{Fe} + x_P \ln x_P), \qquad (2)$$

where $x_{Fe}$ and $x_P$ are the atomic concentrations of Fe and P atoms, respectively. The mixing enthalpy change can be expressed as

$$\Delta H_{mix} = H_{mix} - H_{Fe} x_{Fe} - H_P x_P, \qquad (3)$$

where $H_{mix}$, $H_{Fe}$ and $H_P$ are the enthalpy of mixed phase, pure Fe and pure P, respectively. Then, the second derivative of $\Delta G_{mix}$ with respect to $x_P$ gives:

$$\frac{d^2 \Delta G_{mix}}{dx_P^2} = RT \left[ \frac{1}{x_P} + \frac{1}{1-x_P} \right] + \frac{d^2 H_{mix}}{dx_P^2}. \qquad (4)$$

We calculated $H_{mix}$ through the MD simulations using the FeP potential. Figure 1(b) shows the fitting of $H_{mix}$ as a function of $x_P$ to a second order polynomial. Then, by equating the second derivative of $\Delta G_{mix}$ in Eq.(4) to zero, the spinodal temperature for mixture $Fe_{85}P_{15}$ can be found to be 1985 K, which is far above the glass transition temperature of about 950K [24, 25]. This thermodynamic analysis would predict spinodal decomposition at the glass transition temperature, which is consistent with our observations of the P aggregation in the simulations.

To further examine that compositional segregation plays a strong role in the microstructure, we have created a series of systems annealed at different temperatures as described above. To analyze the phase separation, the $Fe_{85}P_{15}$ system was divided into cubic voxels with side length 5Å, and the P concentrations over these grids were collected. Figure 1(c) shows the difference between the A1 system and other systems, which indicates the degree of segregation as the composition systematically diverges from the mean composition to enhance higher and lower composition regions, the distribution eventually becoming bimodal. Another measure we used to characterize the phase separation is the degree of compositional segregation, $S_P$. We define the $S_P$ as the average number of P atoms within a near-neighbor distance of any P atom as the inset shows in Figure 1(d). The near neighbor distance 3.475Å was chosen from the first trough in the radial distribution function of pure P in a bcc structure modeled with this potential. Although the $S_P$ increases with the annealing temperature, it is always smaller than that found in the MQ system.

During the loading process, the system undergoes failure by the propagation of a crack, consistent with previously reported results [19,20]. Figure 2(a) shows the related stress-strain curves. It can be found that the $Fe_{85}P_{15}$ MGs experience a gradual transition from ductility to brittleness with the increase of annealing temperature. For instance, the A1 sample that failed at ~6.5 GPa exhibited larger plastic strain, 4.3%, as compared to near zero plastic strain in the A800 sample before fracture. During the fracture process, the stress decreases as the strain increases. Every dramatic drop in stress is correlated with the propagation of the crack and every plateau is correlated with a period of plastic deformation in the crack tip process zone [26]. We note that the strain at which the systems completely fracture also increases with the annealing temperature, further indicating a decrease in ductility. We can compute the toughness of these systems in order to quantify their ability to resist fracture [27]. The toughness is defined as the area under the stress-strain curve. While the toughness measured in this manner is not a material property, it is a reasonable measure of the fracture resistance of the material under the specific load conditions given the geometry of the fracture test applied. Our results, given in Figure 2(b) show that the toughness monotonically decreases with the annealing temperature.

Generally speaking, composition plays a key role in determining the microstructure and properties of materials. We can now directly correlate the $S_P$ in this model with the toughness of systems with different compositions that have been exposed to different annealing temperatures. As shown in Figure 3, a similar trend has been found in the $Fe_{80}P_{20}$ and $Fe_{75}P_{25}$ systems as in the $Fe_{85}P_{15}$ system. Moreover, we can expect that an increase in the P concentration will result in a higher $S_P$ at the same annealing temperature. In addition, the system containing more P atoms always has a lower toughness as shown in Figure 3(b).

As mentioned above, the annealing process and the alloy composition affect both the microstructure and the resulting toughness within this model of FeP MGs. Next we check whether there exists any correlation between the $S_P$ and the toughness. Figure 3(c) shows the change of toughness as a function of $S_P$ for all the systems investigated above. It is found that the toughness decreases monotonically with $S_P$ except for the highly unstable A1 samples. For the other well annealed systems, the toughness decreases linearly with $S_P$ independent of chemical composition, as shown in Figure 3(d).

The results above clearly show that changing annealing temperature induces changes in the glass microstructure of this model system arising from phase separation and that this compositional segregation correlates strongly with embrittlement. Previous experimental evidence has shown that processing effects such as cooling rate and processing environment also influence the fracture behavior of MGs [28]. Therefore, the observed phase separation may not be the only or even the primary cause of embrittlement arising from changing processing condition. To explore the fracture mechanism in more detail, Figure 4(a) shows the structural evolution of crack during the tensile process. It can be found that the sharp crack tip propagates towards the P-rich phase which is amenable to nanoscale cavitation. Every sharp spread of crack begins with the expansion of a cavity. The initial fracture surface is introduced as a mathematically precise plane and has a composition that reflects the host alloy. As shown in Figure 4(b), the P concentration increases steadily with the crack propagation, exceeding the average concentration of 15% even at the beginning of fracture and rising to nearly 24%. This indicates that the atomically sharp crack tip propagates along a surface with higher P concentration. This provides strong evidence that the previously described compositional segregation is strongly influencing the fracture behavior of this FeP MG model system.

According to previous studies, MGs containing two glassy phases have remarkable plasticity [29-31]. However, we come to completely opposite conclusions in this work regarding the system at hand. The reason for this puzzling behavior is that the P-rich phase is extremely brittle while the Fe-rich phase is more ductile. Clusters consisting of P atoms have a strong propensity for nanoscale cavitation in the region ahead of the crack tip. This along with the strong spinodal decomposition we observe explains why the atomically sharp crack tip propagates along a fracture plane with higher P concentrations as shown in Figure 4(b).

## 4. Conclusion

In this work, we have investigated the fracture behavior of a particular interatomic potential that has been extensively used to model the fracture behavior of FeP MGs using MD simulations. Our results clearly show that the system is unstable to spinodal decomposition, and due to this fact changing the annealing temperature results in drastically different degrees of compositional segregation. This compositional segregation is the dominant factor determining the resulting toughness of the model MGs. For well-annealed systems, the toughness decreases linearly with the degree of compositional segregation independent of chemical composition. As a result of this investigation we believe that this FeP EAM model is not a realistic system for the investigation of fracture in MG materials. The development of more realistic force fields that exhibit a range of ductility and that do not suffer from such a strong instability to spinodal decomposition would be an excellent goal for future work.

**Acknowledgments**

MLF acknowledges support by NSF under Award No. 1408685/1409560. We also acknowledge the prior work and comments of Prof. Pengfei Guan of the Beijing Computational Sciences Research Center who helped inspire this work and who provided insightful comments on the manuscript.

**Figure Captions**

**Figure 1** (a) Snapshots of the representative initial and final structures of $Fe_{85}P_{15}$ during the traditional melting-and-quenching process. Blue: P atoms; Red: Fe atoms. (b) Change of $H_{mix}$ versus $x_p$, at different temperatures. (c) Fluctuation of P concentrations within 5Å voxels relative to A1 system. (d) $S_P$ as a function of annealing temperature.

**Figure 2** Stress-strain curves (a) and related toughness (b) of MQ-$Fe_{85}P_{15}$ obtained at different annealing temperatures.

**Figure 3** Effect of annealing temperature on $S_P$ (a) and the toughness (b) of different alloy systems. (c) Toughness as a function of $S_P$; (d) Toughness versus $S_P$ of all except the A1 systems.

**Figure 4** Structural evolution of crack during the uniaxial tensile loading of MQ-$Fe_{85}P_{15}$. (a) Snapshots of the crack in a slice with a thickness of 5Å. (b) Crack propagation and the change of $x_P$ along the fracture surface during mechanical loading.

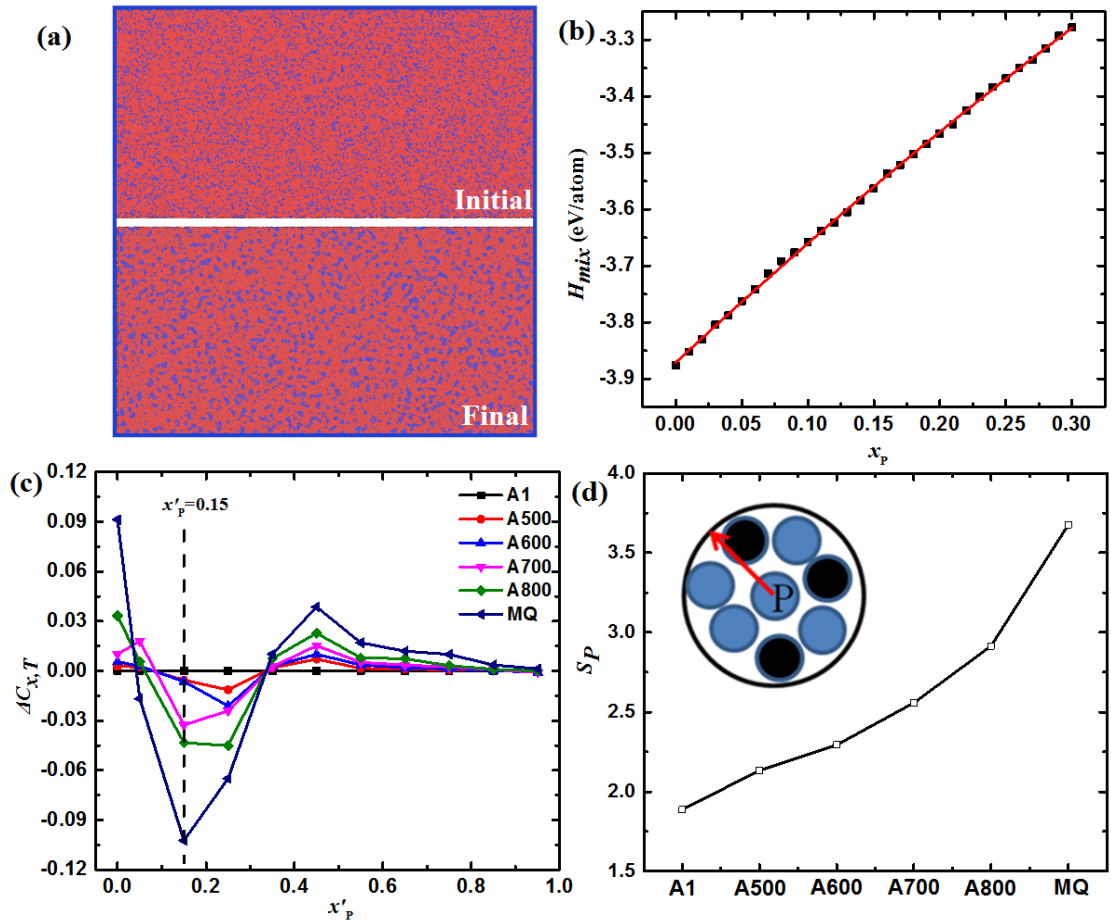

Figure 1

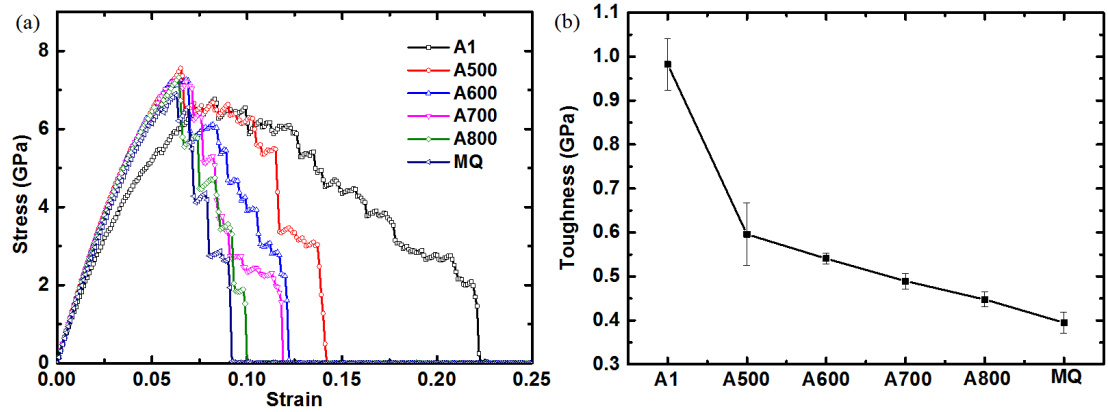

Figure 2

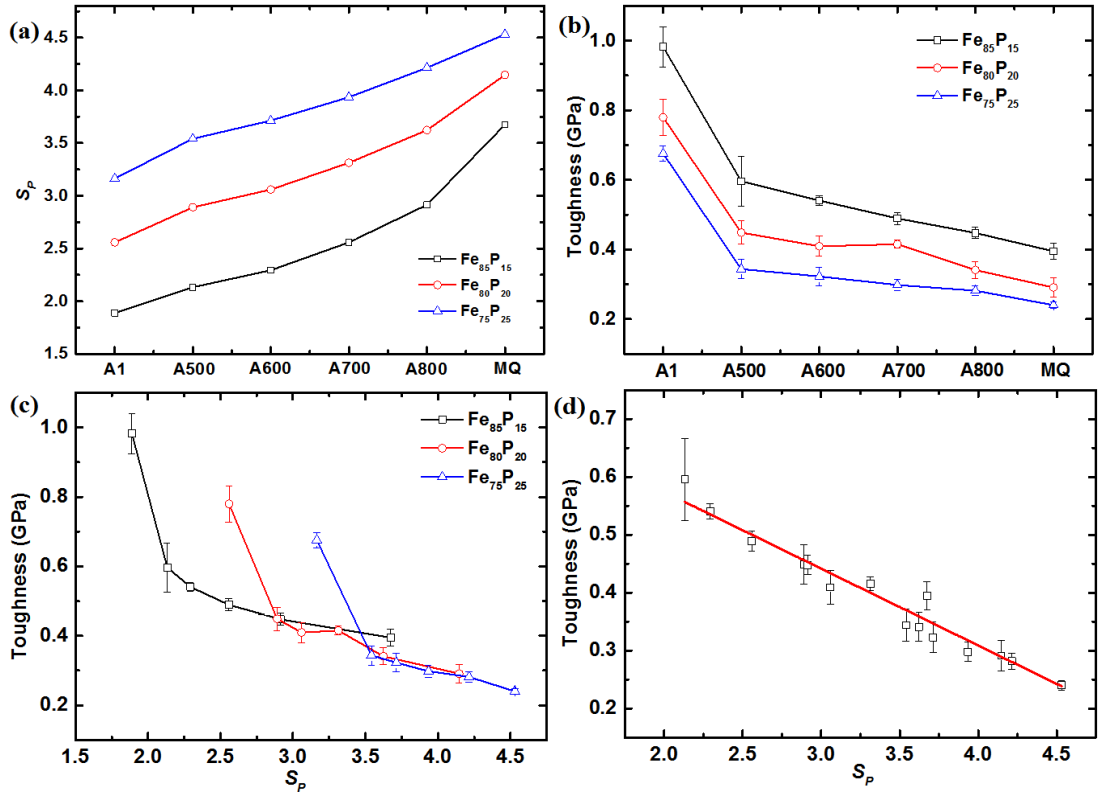

Figure 3

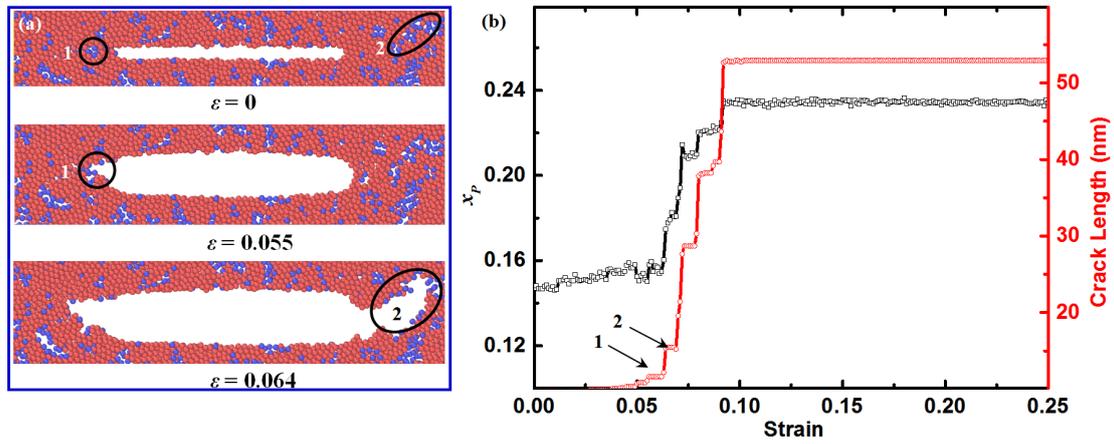

Figure 4